# A Plantar-pressure Based Tongue-placed Tactile Biofeedback System for Balance Improvement

N. VUILLERME*, O. CHENU, N. PINSAULT, A. FLEURY, J. DEMONGEOT and Y. PAYAN

Laboratoire TIMC-IMAG, UMR UJF CNRS 5525, La Tronche, France

Correspondence           *Email: nicolas.vuillerme@imag.fr

## 1 Introduction

Maintaining an upright stance represents a complex task, which is achieved by integrating sensory information from the visual, vestibular and somatosensory systems. When one of these sensory inputs becomes unavailable and/or inaccurate and/or unreliable, postural control generally is degraded. One way to solve this problem is to supplement and/or substitute limited/altered/missing sensory information by providing additional sensory information to the central nervous system via an alternative sensory modality. Along these lines, we developed an original biofeedback system [1] whose underlying principle consists in supplying the user with supplementary sensory information related to foot sole pressure distribution through a tongue-placed output device (Tongue Display Unit, "TDU" [2]). The purpose of the present experiment was to assess its effectiveness in improving balance in young healthy adults.

## 2 Materials and Methods

Ten young healthy adults were asked to stand as immobile as possible with their eyes closed in two conditions of No-Biofeedback and Biofeedback. In the Biofeedback condition, subjects performed the postural task using our biofeedback system. A plantar pressure data acquisition system, consisting of a pair of insoles instrumented with an array of 8×16 pressure sensors per insole was used to collect the magnitude of pressure exerted on each left and right foot sole and to calculate the positions of the resultant centre of foot pressure (CoP). CoP data were then fed back in real time to the TDU, comprising a 2D array (1.5 × 1.5 cm) of 36 electrotactile electrodes each with a 1.4 mm diameter, arranged in a 6 × 6 matrix [1;3]. The matrix of electrodes, maintained in close and permanent contact with the front part of the tongue dorsum, was connected to an external electronic device triggering the electrical signals that stimulate the tactile receptors of the tongue via a flat cable passing out of the mouth. The underlying principle of our biofeedback system was to supply subjects with supplementary information about the position of the CoP relative to a predetermined adjustable "dead zone" (DZ) through the TDU. In the present experiment, antero-posterior and medio-lateral bounds of the DZ were set as the standard deviation of subject's CoP displacements recorded for 10 s preceding each trial. A simple and intuitive coding scheme for the TDU, consisting in a "threshold-alarm" type of feedback was used. When the position of the CoP was determined to be within the DZ, no electrical stimulation was provided in any of the electrodes of the matrix. When the position of the CoP was determined to be outside the DZ, electrical stimulation was provided in distinct zones of the matrix, depending on the position of the CoP relative to the DZ. Specifically, eight different zones located in the front, rear, left, right, front-left, front-right, rear-left, rear-right of the matrix were defined; the activated zone of the matrix corresponded to the position of the CoP relative to the DZ. Centre of foot pressure, recorded using a force platform (AMTI), were processed through a space-time-domain analysis and modeled as fractional Brownian motions according to the procedure of stabilogram-diffusion analysis (SDA) [4].

## 3 Results and Discussion

Space-time domain analysis showed decreased CoP displacements in the Biofeedback relative to No-biofeedback condition[1]. SDA further showed decreased spatio-temporal threshold at which corrective mechanisms are called into play and an increased degree of control of the CoP displacements in the Biofeedback relative to No-biofeedback condition. On the whole, these results suggest that the effectiveness of the biofeedback in decreasing the CoP displacements during upright quiet standing stems from an increased contribution and efficiency of anti-persistent mechanisms involved in the regulation of the CoP.

# 4 Conclusion

The present findings evidence that this new plantar-pressure based tongue-placed tactile biofeedback system can be used to improve balance. An investigation involving individuals with somatosensory loss in the feet from diabetic peripheral neuropathy currently is being conducted to strengthen the potential clinical value of the approach reported in the present experiment.

# References

[1] Vuillerme, N., Chenu, O., Demongeot, J. and Payan, Y., 2007, Controlling posture using a plantar pressure-based, tongue-placed tactile biofeedback system. *Experimental Brain Research*, (DOI 10.1007/s00221-006-0800-4).

[2] Bach-y-Rita, P., Kaczmarek, K.A., Tyler, M.E. and Garcia-Lara, J., 1998, Form perception with a 49-point electrotactile stimulus array on the tongue. *Journal of Rehabilitation Research and Development*, **35**: 427–430.

[3] Vuillerme, N., Chenu, O., Demongeot, J. and Payan, Y., 2006, Improving human ankle joint position sense using an artificial tongue-placed tactile biofeedback. *Neuroscience Letters,* **405**: 19–23.

[4] Collins, J.J. and De Luca, C.J., 1993, Open-loop and closed-loop control of posture: a random-walk analysis of centre-of-pressure trajectories. *Experimental Brain Research,* **95**: 308–318.